# Bright Water

## Hydrosols, Water Conservation and Climate Change

Russell Seitz
Department of Physics  Harvard University
17 Oxford Street, Cambridge MA USA 02138
Seitz@physics.harvard.edu

**Abstract**

Since air-water and water-air interfaces are equally refractive, cloud droplets and microbubbles dispersed in bodies of water reflect sunlight in much the same way. The lifetime of sunlight-reflecting microbubbles, and hence the scale on which they may be applied, depends on Stokes Law and the influence of ambient or added surfactants. Small bubbles backscatter light more efficiently than large ones, opening the possibility of using highly dilute micron-radius hydrosols to substantially brighten surface waters. Such microbubbles can noticeably increase water surface reflectivity, even at volume fractions of parts per million and such loadings can be created at an energy cost as low as J m$^{-2}$ to initiate and milliwatts m$^{-2}$ to sustain. Increasing water albedo in this way can reduce solar energy absorption by as much as 100 W m$^{-2}$, potentially reducing equilibrium temperatures of standing water bodies by several Kelvins. While aerosols injected into the stratosphere tend to alter climate globally, hydrosols can be used to modulate surface albedo, locally and reversibly, without risk of degrading the ozone layer or altering the color of the sky. The low energy cost of microbubbles suggests a new approach to solar radiation management in water conservation and geoengineering: Don't dim the Sun; Brighten the water.

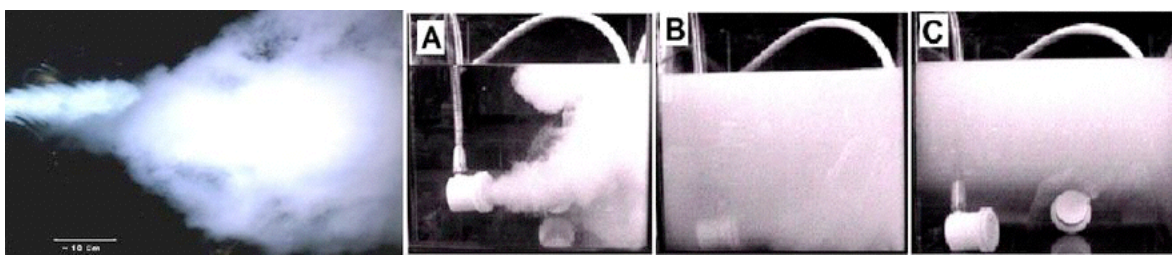

**Figure 1:** Photographs showing injection of ~100 cm$^3$/sec of ~ 1 volume % , ~ 1 micron bubble radius hydrosol into a pool of water (the scale indicates 10 cm). Left photo shows top view during injection. Photographs A and B show side views 10 seconds after initiation and at 60 seconds when injection is stopped. Photograph C shows distribution 120 seconds after end of injection.

**Introduction**

Although the Earth has been described as "a pale blue dot," its albedo is low because the deep water covering two-thirds of its surface is as dark as a blacktop parking lot. The absorption of ~93% of incident solar energy by waters fresh and salt tempers winter temperatures and drives the global hydrologic cycle. As the hydrosphere also stores most of the energy trapped by greenhouse gases, increasing surface albedo to reduce heat oceanic uptake is a potentially important way to counterbalance human-induced warming from landscape darkening and emission of black carbon and greenhouse gases.

Boosting albedo to mitigate microclimates may antedate the edict of Solon that turned Athens into the original "shining city on a hill." The ancient use of whitewash gave way to modern studies of white roof effects half a century ago (Neiberger 1957), but the potential of white water for climate mitigation has been largely overlooked. While air may seem an improbable watercolor pigment its effects are in plain sight -- it provides enough refractive index contrast to brighten ship wakes, waterfalls, and breaking waves. Besides being >$10^3$ less dense than solid pigments, air is free, and microbubbles are already used in paint to extend the covering power of $1000 a tonne titanium dioxide.

Bubbles require relatively little energy to create (Seitz 1958), and can increase the reflected (outgoing) solar flux $F$ from a given body of water by providing voids that backscatter light. As spherical voids in the water column present the same refractive index contrast as round aerosol water droplets suspended in air, hydrosols are essentially atmospheric clouds turned inside out, and the formalism used to express the degree of planetary brightening $\Delta F$, from clouds in the sky can be applied as well to hydrosol clouds in the sea, or any other body of water

$$\Delta F = \Delta R_c \, F_o \, \mu_o \, (1-A_c) \, T^u_c \, T^d_c \tag{1}$$

where $\Delta R_c$ is the change in the albedo of the upper water column (i.e., of the water surface) that bubbles produce, $F_o$ is the solar irradiance, $\mu_o$ is the cosine of the solar zenith angle, $A_c$ is the cloud cover fraction, and $T^u_c$ and $T^d_c$ are the up- and down-welling transmissivities of the air-water column under clear sky conditions. When subsurface bubbles approach aerosol droplets in number density and backscattering efficiency, they can render a water surface almost as bright as clouds in the sky. Though colorless, reflective bubbles increase the surface radiance and change the color of the ocean in ways reflecting the spectral backscattering and absorption of the undisturbed background waters (see figure 4 and Zhang et al. 2004). Unlike plankton blooms that can increase near surface energy absorption and water temperature, augmentation of 'undershine' by microbubbles can reduce net energy absorption. and lead to cooler daytime waters.

This paper considers issues relating to the creation and persistence of microbubbles and the potential for using reflective microbubble dispersions, , to manage solar radiation uptake by locally brightening some of the > 300 million square kilometers of fresh and salt water that cover most of the Earth. Such a dispersion is termed a' *hydrosol' when, as with aerosol particles, its voids are small enough to form a relatively stable suspension.* Potential applications of hydrosols include future solar radiation management on regional to global scales and present water conservation on the scale of lakes, reservoirs, cooling ponds, and canals.

**Hydrosol Production and Lifetime**

Hydrosols are readily produced by methods involving expansion of air saturated water through vortex nozzles (See figure 1) or shear generating constrictions, and by the use of mechanical shakers or ultrasonic transducers that create local Weber numbers higher than for typical water surface waves While visible whitecap bubbles typically rise and burst in seconds, microbubbles take far longer to surface than the smallest the naked eye can see. That this is the case is evident from Stoke's Law

$$\mathbf{V}_{St} = 2\ \mathbf{g}\ \mathbf{r}^2 / 9\ \nu \tag{2}$$

where in the rigid sphere approximation, $\nu \propto \eta / \rho$, $\eta$ is the dynamic viscosity and $\rho$ is the fluid density.

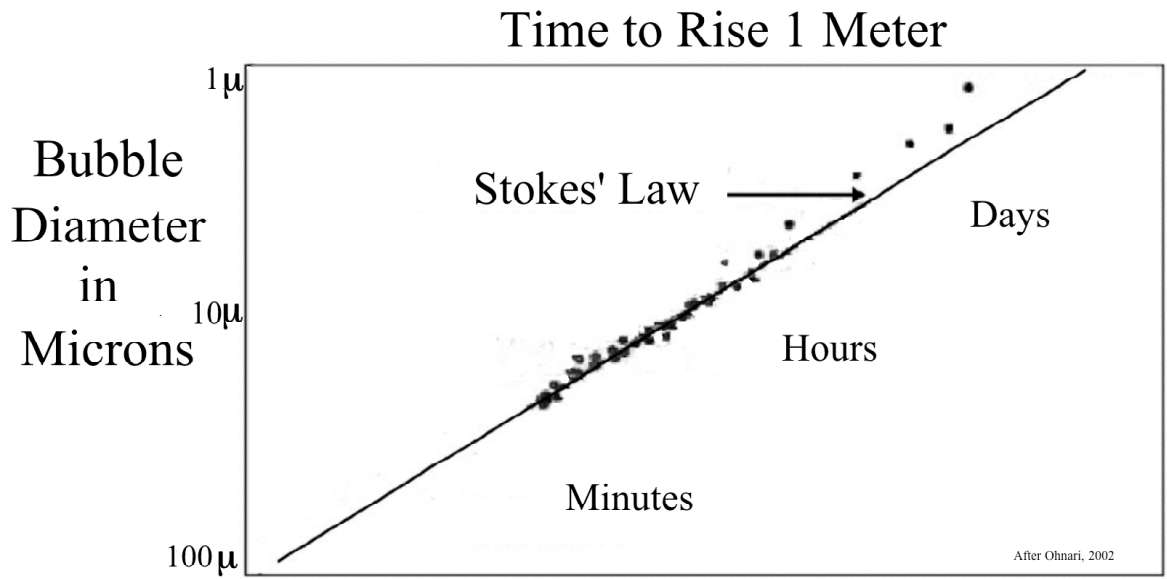

**Figure 2: The rate of rise of microbubbles in still water as a function of bubble diameter based on studies of microbubble populations generated by a single stage, Riverforest Corporation AS-series vortex flow nanobubble generator**

This velocity-viscosity relationship limits the stagnation (*St*) velocity of microbubbles to millimeters per hour rather than centimeters per second, and in consequence just as aerosol droplets are too small to fall rapidly through the air, hydrosol microbubbles are too small to rise rapidly through water, leading in both situations to relatively long lasting reflective clouds or light scattering haze. Figure 2 compares experimental data from Onhari et al. (2002) with theoretical estimates for the time for a bubble of a particular size to rise one meter in still water.

Because gas solubilities in water differ, bubble shrinkage results in nitrogen, argon, and heavy isotope enrichment in residual gas within bubbles, and oxygen and $CO_2$ depletion. In rough water, near surface nitrogen supersaturation may extend microbubble lifetime, and microbubble injection may itself supersaturate small reservoirs or thin surface layers in the ocean.

Microbubble lifetime is further complicated by the variable concentration of surfactants in natural waters and man-made water bodies such as aqueducts and reservoirs. In the decades following sonar researcher Alexander Hiller's 1967 discovery of a stable population of micron-sized bubbles in seawater (B. Johnson, personal communication), oceanographers postulated the molecular self-assembly of surfactant films at the air-water interface and the stabilization of microbubbles by colloids. As compared to short (minutes) expected lifetimes in pure water, analyses found that silane surfactants increased microbubble lifetimes to hours or days (Johnson and Cooke 1981). Thorpe (1982) and Johnson and Wangersky (1987) found that over a yield range from 2% to 93%, the yield of stable microbubbles in coastal seawater is correlated to surfactant and colloid concentrations.

That such factors can affect bubble lifetime has also been widely observed under realistic conditions. For example, studies of sonar reflectance by Weber et al. (2005) indicated seawater microbubbles produced in an oceanographic research vessel's wake lasted ~15 times longer than theory predicted for surfactant-free water, and Lozano et al. (2007) observed a doubling of sonified seawater microbubble lifetime commensurate with monolayer formation. While there are indications that bubble lifetime in natural waters can be much longer than theory suggests for pure water in vitro (Lozano and Longo 2009), little more than occasional observations is available to provide information about the seasonal and geographic variation of ocean microbubble persistence (Johnson and Cooke 1981).

In addition to chemical surfactant effects, physical mechanisms can contribute to microbubble stabilization (Abkarian et al. 2007). Seawater typically contains ~$10^7$ virus-like nanoparticles per $cm^3$, and it has been demonstrated that microbubbles can be stabilized by the physical accretion of such micro- or nanoparticles at the air–water interface to form polygonal domains (Dressaire et al, 2008). Organosilicon compounds also promote such effects and persistent inert gas hydrosols have been developed for use as nuclear-magnetic resonance (NMR) contrast agents (Qin et al. 2009). On a time scale of minutes, molecules of the biological surfactants oceanographers term *gelbstoff* accumulate into patchy Langmuir-Blodget films that interact with ambient nanoparticles (e.g., viral capsids, diatom fragments, and debris from plankton decay and digestion) to coat the air-water interface (Zhang et al. 1998).

As a result, seawater is less a salt solution than a thin colloidal soup, and the water surface often hosts a microbial community, or *pneumonueston*, that enriches the surface layer in gelatinous polysaccharide and protein surfactants that impact bubble nucleation and persistence (Wurl and Holmes 2008). The longer that bubbles rise or move through biologically active waters, the greater their accumulation of ambient surfactants and nanoparticles. Prolonging these processes may promote the formation of virtual micelles (Pu et al. 2006), the surface energy of which retards the kinetics of gas dissolution into the surrounding fluid  As bubbles lose gas and shrink, lipids and long chain molecules are further  concentrated at the air-water interface.

Marine algal derivatives like carrageen are used extensively to stabilize microbubbles in food products, and can form liquid crystal layers that extend bubble lifetimes to months or years (Dressaire et al. 2008). Many vacuole containing algae and diatoms (e.g., *microcystis* and *synechococcus*), and white calcite coated coccolithophores are highly reflective, (Gargett 1991) and it is possible climate feedback loops involving albedophoric phytoplankton may have contributed to a sort of geoengineering by evolutionary misadventure in albedo-driven palaeoclimate events like the 'snowball Earth' episodes (Hoffman et al. 1998). While the practical limits of microbubble persistence remain to be discovered, what appear to be the remains of bubble-coating organic films appear in lake sediment microfossils.

The longer the lifetime of the microbubbles, the larger the scale on which they can be practically deployed. Natural wave breaking mainly produces ephemeral macrobubbles, as relatively low Weber numbers in whitecaps rarely give rise to jetting and high velocity air entrainment. In contrast, artificial two-phase flows can be optimized to efficiently produce narrow microbubble size distributions. The formation of stable interfacial films can also be promoted by lowering the ambient fluid pressure to transiently expand bubbles to create a larger air-water interface at which surfactant molecules can collect, forming robust films when the bubbles contract.

While the rise rate of the bubbles is also important in determining bubble lifetime, vertical-mixing velocities in open bodies of water often exceed $\mathbf{V}_{St}$ for microbubbles. As a result, hydrosols released near the water surface may circulate and diffuse through mixing driven by wind and convection, and microbubbles in quasi-Brownian motion may accordingly be transported downwards and disappear at rates determined more by interfacial gas solution kinetics than by surface bursting.

**The Radiative Effects of Microbubbles**

Compared to the ocean surface albedo of about 0.05-0.10 for calm seas, the bubbles created by ocean whitecaps create an effective reflectance of 0.22 (Moore et al. 2000). At ordinary marine wind speeds (i.e., ~4-8 m/sec), however, whitecaps cover too little area to significantly raise ocean albedo. While storm force winds can produce more noticeable effects such conditions are limited in time and space (Willis 1971; Whitlock et al. 1982), but understanding natural bubble effects may afford insight into the potential of more reflective artificial hydrosols to increase the ocean albedo as a means to cool the planet.

Because light backscattering is cross-section rather than mass or volume dependent, microbubble hydrosols are optically analogous to tropospheric droplet clouds or stratospheric aerosols. Like surface whitecaps, they are white because the underlying water column approximates to a succession of thin slabs. Each thin vertical layer contains few bubbles and intercepts little incident light, but over several meters depth the ensemble can present a projected area as great as the water surface, producing a high bulk backscattering coefficient $B_b$, and a large return of incident energy (Monahan and MacNiocaill 1986; Piskozub et al. 2009). While white pigment particles absorb light internally via intrinsic electronic transitions and band gap phenomena, bubble backscattering is largely attenuated by absorption in the surrounding water.

While clouds in the atmosphere contain ~$10^7$ to $10^9$ droplets m$^{-3}$, seawater generally contains from $10^4$ to $10^7$ microbubbles m$^{-3}$, with a combined volume of some cubic millimeters m$^{-3}$. Optical buoy measurements and hyperspectral satellite observations both confirm that ambient microbubbles do measurably alter the ocean's return of solar energy to space (Terrill et al. 2001; Stramski and Tegowski 2001; Jin et al. 2002). Although the relatively coarse natural microbubbles (diameter ~10-100 microns) in near surface seawater comprise only ~$10^{-6}$ to $10^{-7}$ by volume, Zhang et al. (1998) calculated that the spectral sensitivity of backscattering from an ambient population of $10^6$ ocean microbubbles m$^{-3}$ with diameters in the 10-140 micron range. Because the bubbles in natural surface waters typically occupy a volume fraction of air of only a part per million or less and their collective backscattering cross-section is small, the "undershine" they contribute ordinarily increases Earth's albedo by only ~$10^{-3}$ to $10^{-4}$.

However, solar energy reflection rises dramatically with decreasing microbubble size. Dividing the same volume fraction of air among a larger number of smaller bubbles can cause an exponential rise in a bubble cloud's backscattering cross-section because the air in a single 1 cm$^3$ bubble will fill a trillion one micron ones--. Thus, while one ppmv of 100-micron bubbles may present a scattering cross section of only a few cm$^2$ and reflect only milliwatts m$^{-2}$, the cross section of one ppmv of 1-micron bubbles is three orders of magnitude larger. Because photons penetrating the sea surface can encounter several bubbles in the first few meters of path length (Toole et al. 2000), <1 ppmv of micron-sized bubbles can lead to reflection of watts per square meter of solar energy [see Figures 3 and 4

Because microbubbles are almost as refringent as the glass microbeads used in reflective road signs, backscattering from a sufficient number of very small bubbles in the upper water column may raise the albedo of the ocean-atmosphere interface to the level of bright cloud cover. The backscattering coefficient ($b_b$) of hydrosols of micron-sized bubbles depends on the fraction of incident light that is intercepted and returned between 90º and 180º. The light return is a function of sun angle, the projected microbubble area (or number $n$ of bubbles), and the dimensionless Mie scattering efficiency term $Q_{Bb}$. reflects the refractive indices and other optical property dependent variables The equation for the dimensionless bubble backscattering efficiency coefficient $b_b$ is

$$b_b = \int_{r_{min}}^{r_{max}} Q\, b_b(r)\, \pi\, r^2\, n(r)\, dr \qquad (3)$$

This relationship yields a backscattering cross-section, σ, that is inversely proportional to bubble radius. This integral, which can be used to correct ocean albedo for undershine, has been incorporated into model parameterizations operating over the ranges of bubble population density, size, vertical distribution, and seawater clarity encountered in oceanography to produce the relationships used to calibrate remote measurements of biomass productivity.

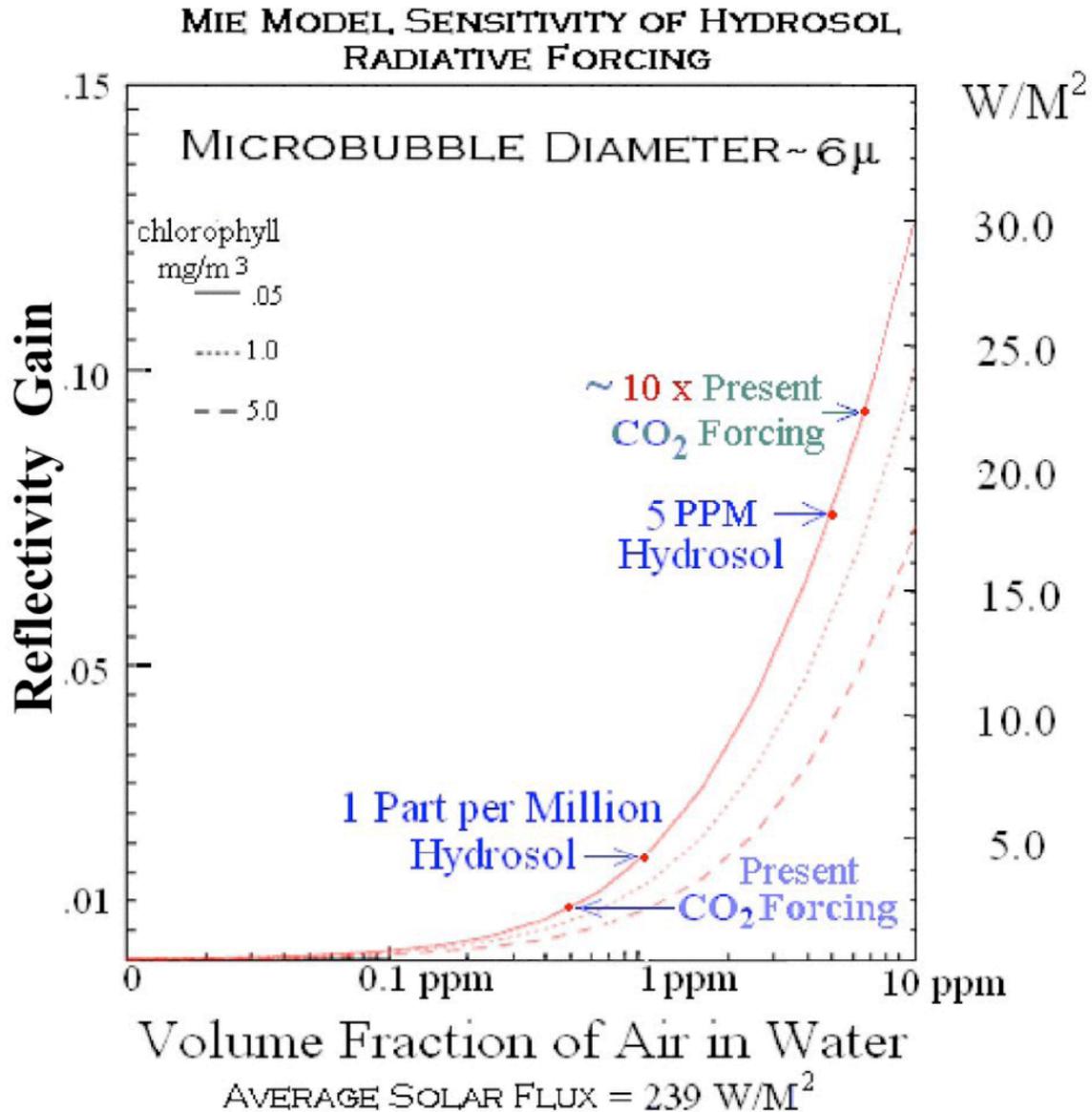

**Figure 3** Mie model results ( see Jin et al 2006 for model details ) indicating the increase in the reflectivity gain (i.e., albedo increase) that can be achieved as a result of increasing the volume fraction of air in water using 6 micron bubble diameter. Results are shown for differing amounts of chlorophyll in the water, and for various $CO_2$ forcing levels

Jin et al. (2002) have used this type of model to calculate the broadband (UVB to 4 micron) albedo shift arising from a natural hydrosol of 10 to 150 μ bubbles. In the calculation, the downward solar flux at the surface was set to 239 W/m² to approximate the average of the effects of the Earth's rotation and cloud cover. The model incorporates absorption due to chlorophyll in seawater as well as light attenuation by the atmosphere and hydrosphere, but neglects spectral variation of reflectivity with water roughness. Initial extrapolation of their results and those of Zhang et al. (1998) suggest that ocean albedo would rise exponentially for microbubble number densities in excess of ~$2 \times 10^7$ m$^{-3}$, or about 1 part per million by volume, and yield albedo increases as high as ~0.012 at a hydrosol concentration of ~ 5 ppmv by volume.
it economically feasible (e.g., over coral reefs) to lower peak water temperatures for periods of months or more on a regional scale.

Assuming a mid-range climate sensitivity of ~0.7 K/Wm$^{-2}$ (IPCC 2007), these results suggest that a natural hydrosol loading of 1 ppmv over the global ocean would increase the oceanic albedo sufficiently to decrease global average temperature by ~1 K. Were the ambient ocean concentration of these relatively coarse (i.e., 10 to 150 μ diameter) microbubbles to exceed 5 ppmv, the increase in the ocean albedo could offset the present radiative forcing from the human-caused increases in the concentrations of $CO_2$, $CH_4$, $N_2O$, and halocarbons.

Were the average bubble diameter in the 0.5 to 3-micron range, as is the case for the bubbles shown in the generation process displayed in Figure 1, the albedo change could be induced with a much smaller hydrosol loading. Mie model sensitivity tests by Jin et al. (2006) indicate that a hydrosol loading of ~5 ppmv in this smaller size range could increase the planetary albedo by ~0.10, and that loading of 10 ppmv of bubbles could produce a negative radiative forcing approximating that of the average global cloud cover, which Schneider (1996) estimates as about -13 Wm$^{-2}$. So large an albedo gain from so small a hydrosol concentration suggests that the radiative forcing due to the rising $CO_2$ concentration could be offset by dispersing about one part per billion of the mass of the atmosphere as microbubbles in the ocean.

**Potential Applications of Microbubble Injection**

In this section, we calculate the scale of the injections that would be required for two potential applications of microbubble injection, one focused on reducing evaporation from reservoirs, cooling ponds and other water bodies as a step toward conserving water and improving energy efficiency, and the second on increasing ocean surface albedo in order to reduce planetary absorption of solar radiation. Increased rates of evaporation from increases in average temperature will augment the stresses on freshwater resources caused by climate change; thus, if the energy driving evaporation from such water bodies can be reduced, evaporative water losses can be reduced. Implemented on regional to global scales in ocean surface waters to reduce the absorption of solar radiation by the Earth-atmosphere system, as is proposed as a means for geoengineering the climate, has the potential to limit climate change.

The energy cost of creating a hydrosol of micron-radius bubbles is small because bubble nucleation in supersaturated media is often thermodynamically favored. The initial inflation energy equals the Laplace pressure (**P = 2 γ r**), which, given the surface energy of water, is ~160 kilopascal in micron-radius bubbles and integrates to ~100 J l$^{-1}$ of internal volume, to which must be added the interfacial energy and the viscous and gravitational work of displacement. (Brennen, 1995) In gas-free water, the pressure due to interfacial curvature rapidly drives micron-sized bubbles into solution, but in air-saturated water (22 mg l$^{-1}$ at STP),but in the presence of film-forming impurities, either present naturally or intentionally added, microbubbles may last for tens to hundreds of seconds. Microbubbles have been observed to persist even longer in biologically active seawater, if interfacial effects lead to spontaneous encapsulation, In times and places where hydrosol lifetimes of hours or longer are obtainable, multi-kilometer coverage may be possible. especially over coral reefs and in shallow estuaries, where high natural surfactant concentrations and subsequent hydrosols persistence may make it economically feasible to lower peak water temperatures for extended periods.

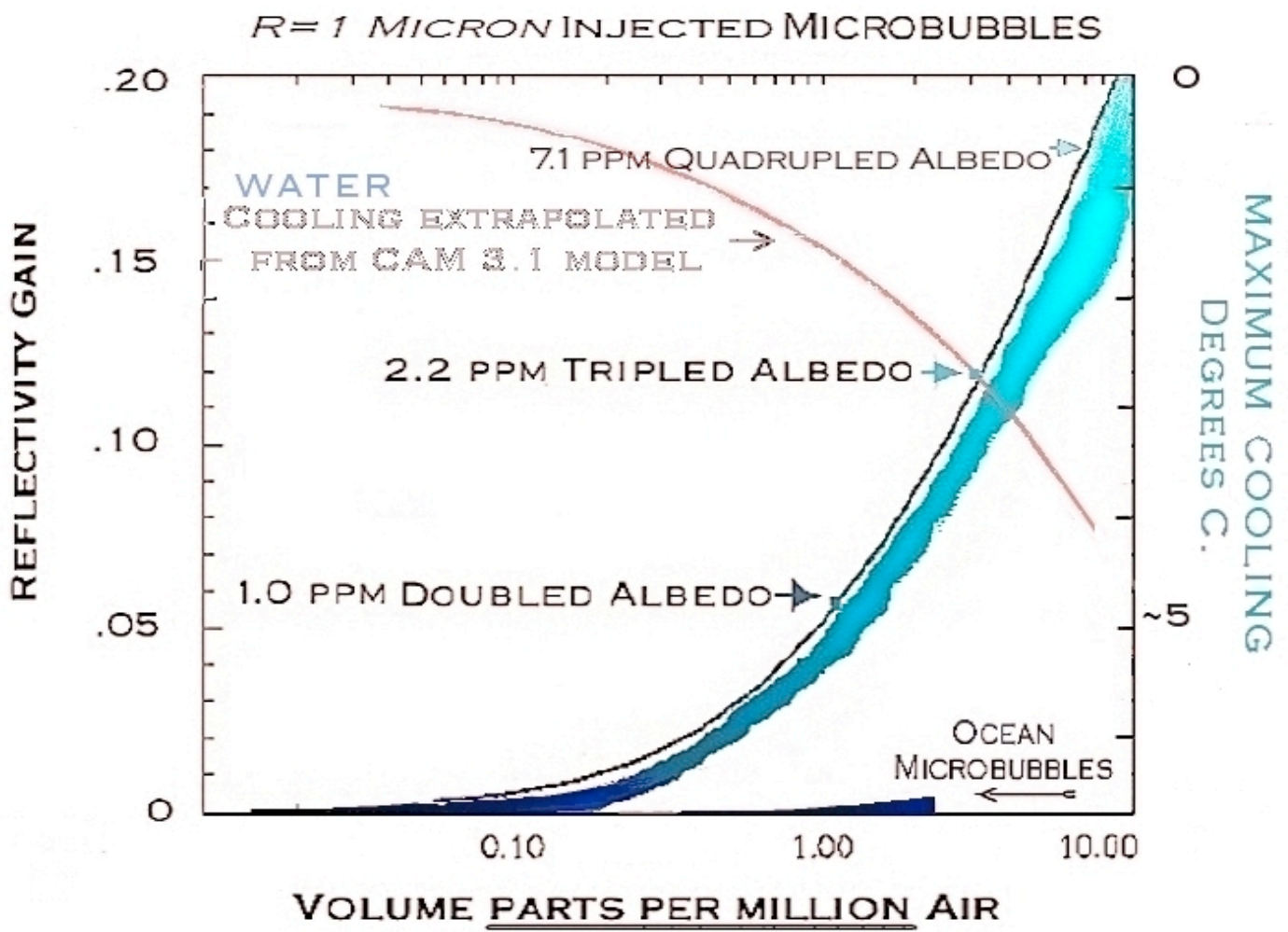

*Figure 4 Reflectivity gain (from Mie model of 1 micron bubbles) versus surface water cooling (from 6 micron bubbles in CAM-1 3-D GCM) ;Color adjacent to reflectivity curve approximates visual shift in water appearance versus microbubble concentration. Inset shows scale of natural microbubble contribution to ocean albedo*

Because natural surfactant concentrations vary widely and with time of year, hydrosol stabilization may at times require pre-concentration of natural surfactants or addition of synthetic, biodegradable ones. Since seawater hydrosols survive dilution with fresh water, salt-water hydrosol concentrates might be piped into fresh water to stem evaporation without synthetic surfactants and with only part per million increases in salinity. While the relationship between local biota and interfacial chemistry is beyond this paper's scope, biologically productive waters appear to extend the lifetime of both marine microbubbles and those produced *in vitro*. The large variation in spontaneous bubble stabilization (Johnson and Cooke 1981), however, makes clear the need for further research to correlate microbubble lifetimes with water composition, natural surfactant availability, and colloidal and organic nanoparticle content.

Because water absorbs red light more strongly than blue and suspended matter attenuates sunlight by $<1$ to $> 5 \cdot 10^{-2}$ $m^{-1}$, shallow hydrosols backscatter more solar energy than deeply immersed ones. Because organic molecules generally raise the refractive index of the interfacial layer, their introduction to extend bubble lifetime has the potential to increase reflection by up to 400% (Zhang et al. 1998).

Commercial hydrosol generators can produce microbubble concentrates of ~80,000 - 100,000 ppmv, which have the potential to double the albedo even when diluted by a factor of $10^3$-$10^4$; It is noteworthy that ships equipped with sub millimeter bubble generators to reduce hull drag, are under construction (Kato 1999). Because common biodegradable foaming agents stabilize bubbles at mass concentrations of 10-100 ppm, surfactant concentrations following bulk hydrosol dilution would be in the part per billion range (~mg $m^{-3}$), approximating the initial mass of air in the microbubbles themselves. Under such conditions, the cost of surfactants (Hosseinet al. 2010) to achieve a measure of artificial microbubble stabilization would be <$10 per kilogram, or < $100 $km^{-1}$. Although the surfactant-induced viscosity may increase the energy cost of making microbubbles, the low internal pressures (~$10^2$ kilopascal) involved suggest that only megajoules $km^{-2}$ might be needed to transiently reduce surface temperatures by 1 K or more (R. Garwin, Personal communication)

To calculate the potential effects of microbubble injection, the CAM 3.1 global circulation model developed at the National Center for Atmospheric Research has been used to simulate the climate's response to an increase in global ocean albedo. Baseline simulations were conducted the $CO_2$ concentration at preindustrial (280 ppmv), current (390 ppmv), and double current (780 ppmv) levels. Using the doubled $CO_2$ case (780 ppmv) as the control, hydrosol simulations were conducted to examine the counter-balancing effects of increasing ocean albedo by 0.01 and 0.05 above its present average value of ~0.06. For reference, increasing the surface albedo by 0.05 would require an air concentration of ~$2.5 \times 10^{-6}$ by volume and a microbubble diameter of 6 microns; alternatively, this approximates the albedo increase that would result from an upper water column void fraction of $10^{-8}$ of 1-micron bubbles.

Seasonal cycle simulations were run using a 2° by 2.5° latitude-longitude resolution, employing an upper ocean slab model with specified seasonally varying meridional ocean heat transport]. Local sea surface temperature is then determined by the local energy balance established through the coupling to the atmosphere and the specified ocean heat transport. All simulations were carried out for 70 years, with results for the first 40 years discarded in the analyses to assure the simulations have reached a new equilibrium state.

*Figure 5 , courtesy of Kenneth Caldeira & the Carnagie Institute, shows a CAM 3.1 model projection for a $CO_2$ concentration of 780 ppmv and the ocean albedo increased by 0.05., raising global clear sky albedo by ~.031*

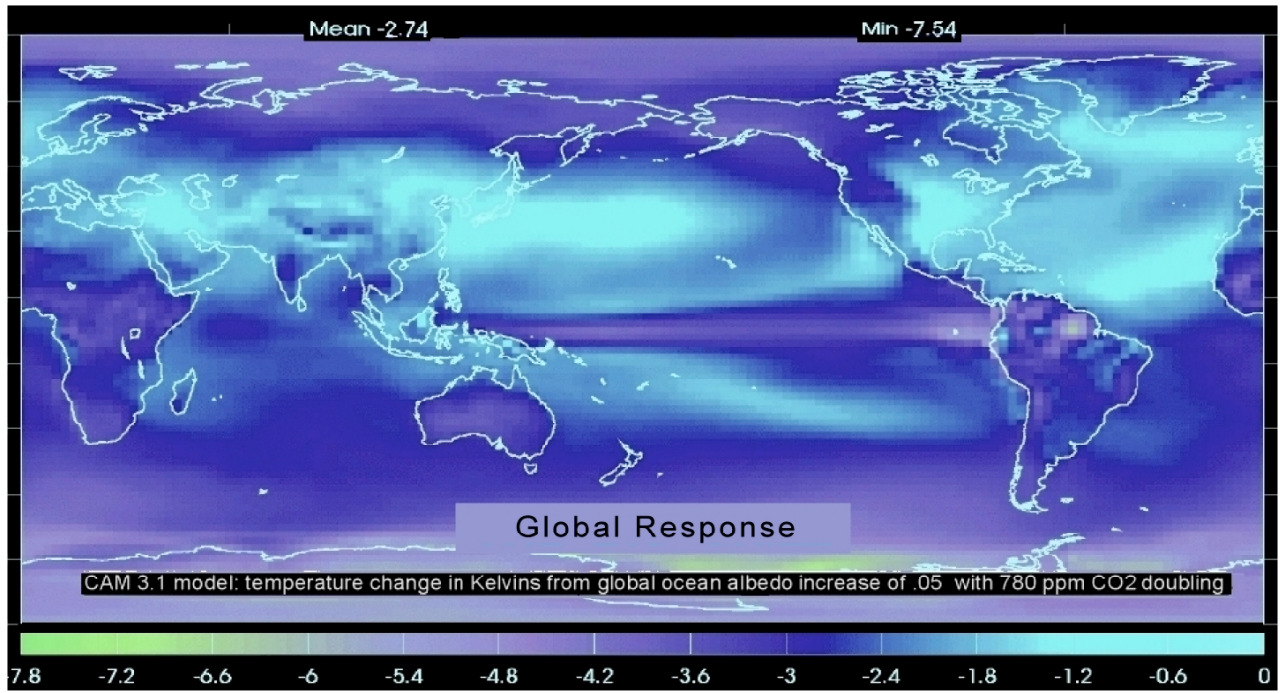

Figure 5 shows the change in temperature for the case with a $CO_2$ concentration of 780 ppmv and the ocean albedo increased by 0.05. With 25% coverage by marine cloud cover, the hydrosols increased the global, top-of-the-atmosphere albedo by ~0.175 over the 70% of the world covered with water. This increases the overall planetary albedo by 0.05, times the ratio of sea to land area, producing a global albedo gain of ~0.031, which in the model runs reduced global average surface temperature by ~ 2.7 K, an overall cooling greater than the warming induced by doubling the baseline $CO_2$ concentration of 390 ppmv to 780 ppmv. Some ocean and continental areas exhibited cooling in excess of 5 K, suggesting that a boost of ocean albedo by 0.05 would be somewhat larger than the change required to offset the warming influence of a $CO_2$ doubling.

Rather than increasing the ocean albedo over the global ocean, creating a larger albedo change in a subset of regions also has the potential to cool the climate—and likely at lower cost. For example, focusing the hydrosol injection over low latitude ocean areas with higher than average insolation would seem to be an obvious step. Such an effect might well be achieved by creating 'virtual icecaps' of bright water in the tropics by deploying equatorial arrays of hydrosol generators that would phase their bubble generation with local peak sun conditions.

Using microbubbles to conserve terrestrial water resources might well be effective and economical for water bodies such as reservoirs and cooling ponds. The analysis of Akbari et al. (2009) estimated that the reduction in top-of-the-atmosphere radiative forcing from increasing the albedo of 1 $m^2$ by 0.01 would offset the warming influence of 2.55 kg of $CO_2$. This suggests that a prolonged doubling of the reflectivity of a hectare of water (e.g., from 0.06 to 0.12), would offset the emission of >150 tonnes of $CO_2$, while providing the important co-benefit of conserving water.

As an example of a specific application, it might well be feasible for fair weather operation of a small pump powered by human labor or a small photoelectric array to inject enough bubbles into a 1-hectare water-holding pond to reduce evaporation by a millimeter per day (out of, for example, a typical rate of a few m $yr^{-1}$). If this could be accomplished, it would save a small farmer 3,600 tonnes of irrigation water per year. Similarly, by reducing evaporation, reservoir brightening would increase production of hydroelectric power, reducing the need for fossil fuel generated power. If hydrosol generation by expansion of compressed air saturated water approached theoretical efficiency, the energy cost of micron sized bubbles would be on the order of a kw-hr $m^{-3}$ of contained air (Ohnari et al. 2006), and megawatt scale wind turbines might produce ~ 2 x $10^4$ $m^{-3}$ of 1-10 micron bubbles a day, which, depending on lifetime, might usefully brighten $10^2$-$10^4$ $km^2$ of surface water. Raising the albedo of rivers might also be used to mitigate power plant thermal pollution, and brightening cooling ponds to reduce solar heating may likewise reduce power plant $CO_2$ emission by increasing thermodynamic efficiency.

**Discussion**

That hydrosols can be used to increase surface albedo and thereby limit Earth system absorption of solar radiation adds this approach to the list of possible approaches for geoengineering the climate (MacCracken 2009). To determine the priority appropriate for study and potential deployment of hydrosols, a comparison is needed of advantages and disadvantages with respect to other approaches, such as stratospheric aerosol deployment (Crutzen 2006; Rasch et al. 2008), augmentation of cloud cover (Latham 1990, 2002; Salter 2008), and direct steps to reduce or reverse increases in the $CO_2$ concentration (Keith et al. 2005).

Compared to use of hydrosols, decreasing planetary absorption of solar radiation by low-latitude injection of stratospheric aerosols reduces the injection and distribution costs because the lifetime of the aerosol particles is roughly two years and the stratospheric circulation widely distributes the aerosols. Reducing surface air temperature by a few K is estimated to require ongoing injection of perhaps 10 TgS/yr (Robock et al. 2009), which would tend to whiten the sky by increasing the natural conversion of direct to diffuse radiation. If the longest seawater microbubble lifetimes reported could be achieved artificially, sustaining a hydrosol concentration of ~500 ppbv (equivalent to ~1 mg $m^{-3}$ of air, or 1 kg $km^{-2}$ $m^{-1}$ over a depth of 10 m), which integrates to a global injection of ~ 50 Tg $yr^{-1}$, would increase the surface albedo over the global ocean by enough to generate a cooling of a few K.

Although these calculations do not account for how reflectivity and surface roughness would evolve during bubble shrinkage and dissolution by limiting the injections to low latitudes during daylight hours, the required microbubble air mass might conceivably be reduced to levels approaching the <10 Tg yr$^{-1}$ mass of $SO_2$ needed to increase stratospheric aerosol loading to counterbalance a $CO_2$ increase, but without the need to loft the mass or alter sunlight quality. Depending on the average hydrosol lifetime, offsetting doubled $CO_2$ might thus require annual injection of only 100- 1000 kg of microbubbles per capita, a level of necessary effort within the limits of industrial precedent, as the mass requirements are smaller in scope than global $CO_2$ emissions. Although such estimates depend on factors such as the climate sensitivity to $CO_2$ doubling, cloud cover and ocean optics uncertainties. Extrapolating from results for laboratory microbubble generators without assuming any benefit of economies of scale (Sadatomi et al. 2007), it appears that some tens to hundreds of gigawatts might suffice to offset petawatts of $CO_2$ induced radiative forcing. This would provide an energy gain of $10^4$ to $10^6$ at a global energy expenditure of only tens of watts of power *per capita*.

As to the practicalities of microbubble generation and dispersion, significant advantage could be taken of the existing fleet of ocean-going vessels, of which >$10^4$ are at sea on any given day. Vortex entrainment of air produces microbubbles more efficiently than wave breaking (Hwang et al. 1989), and shipboard compressors developed to reduce hull drag and fuel consumption by releasing macrobubbles (Kato 1999; Graham-Rove 2008) might also supply microbubble generators to provide hydrosols to amplify wake reflectance at sea (Gordon 1985; Zhang Lewis, Bissett et al. 2004). With conventional wakes extending kilometers astern, the thousands of oil-fueled ships normally underway globally might brighten some $10^5$ to $10^7$ km$^2$ of ocean, to offset both the decrease in planetary albedo produced by the black carbon they emit (Seitz 1991), and the radiative forcing from the ~$10^9$ tonnes of $CO_2$ per year marine transportation currently releases (Buhaug et al. 2009).

As an example of a potential application that would take advantage of the ability to create localized effects, bubble-generating ships might be deployed to lower sea surface temperatures by *en echelon* release of wake microbubbles along tropical storm tracks. Such an approach to reducing tropical cyclone intensity would also likely require less infrastructure, mass, and energy investment than proposals to use wave motion to sink warm surface water that involve gigatonne-scale mass transport (Morton, 2009)

Unlike persistent stratospheric aerosols, hydrosols can be modulated on a time scale of days should unforeseen ecological stresses (Schneider 1996), volcanic eruptions, or extraordinary weather conditions arise. Because hydrosols scatter light forward as well as back into the atmosphere, the underwater light-level reductions considered here are less than biota in these layers encounter under cloudy skies (Johnson and Cooke 1980). In marine photosynthesis largely occurs at depths where light levels are already low, small albedo increases may primarily shift the photosynthetic compensation depth rather than significantly reduce biomass productivity (Dickey and Falkowski 2003). Therefore, as with terrestrial plants (Mercado et al. 2009), productivity may actually increase under diffuse illumination.

As to terrestrial applications, many nations lose more fresh water than they consume (Gokbulak and Ozban 2006). Dimensional analysis suggests dilute hydrosols could reduce solar evaporation from open canals and distribution systems where moving water produces high rates of loss. While efforts to stem evaporation with Langmuir- Blodget monolayers of surfactants like cetyl alcohol have met with only modest economic success because of the high cost of keeping the entire water surface covered, multi-molecular films lining immersed microbubbles are relatively immune to being stranded alee by wind action, and so may prove to be a more cost-effective approach.

**Conclusions**

The most recent IPCC assessment (IPCC 2007) notes that even with an immediate policy response (Van Vuuren et al. 2009), "an average minimum warming of ≈1.4°C (with a full range of 0.5-2.8°C) remains for even the most stringent stabilization scenarios analyzed." Arctic sea ice shrinkage, thawing permafrost, and loss of mass from the Greenland and Antarctic ice sheets raises concern that the climate is approaching conditions that may significantly and adversely influence environment and society. This has elicited prudential calls for research into potential geoengineering options (Shepherd 2009) – and of the hazards that attend them (Victor et al. 2009; Lovelock 2009). Such research is needed to explore the possibility of limiting global warming and its impacts should political efforts to drastically reduce emission levels of greenhouse gases fail, or releases from natural sources accelerate in delayed response to past human action.

While much attention has been focused on increasing the sulfate loading of the stratosphere as an approach to doing this, injection of microbubbles to increase the albedo of the world's oceans and of smaller water bodies appears to be an attractive alternative approach. Hydrosphere albedo already plays a dominant role in determining the global average temperature. Therefore, injecting air into water to adjust its reflectivity may afford fewer ecological surprises or risks to the ozone layer, (Read 2009) than injection of chemical aerosols. In particular, marine photosynthesis is more commonly nutrient rather than sunlight limited, and fresh and salt-water ecosystems have coexisted with wind-generated microbubbles throughout their evolutionary history (D'Arrigo et al. 1984; Katz et al. 2004). While stratospheric aerosol injection (Shepherd 2009; Teller et al. 1997) is aimed at changing global and annual-average conditions and increasing the brightness of clouds by increasing the number of cloud condensation nuclei (Latham et al. 2008) is subject to the vagaries of tropospheric weather and the location of marine stratus clouds, hydrosols can have influences on both the weather and the climate and may offer flexible, local, and reversible modulation of the planetary albedo without injecting materials into the atmosphere.

In addition, bubble injection can assist in both adaptation and mitigation. While analyses indicate that petagram reductions in annual $CO_2$ emissions will be required to stabilize the global climate (IPCC 2007), the per capita carbon emissions of many of the poorest nations are only kilograms per day and their priorities are focused on survival rather than emissions reduction. Because bubble brightening of local waters has the potential to reduce evaporation and thereby increase water resources, it seems likely to become an important technology for adapting to climate change. For example, the low energy cost and void fractions of air needed to increase reflectivity on hectare scales make even short-lived hydrosols of obvious use in water conservation, and brightening water bodies would also reduce net radiative forcing (sustained brightening of 100 m$^2$ roughly offsets emission of one tonne of $CO_2$) allowing arid nations to simultaneously participate in domestic water conservation and international climate stabilization efforts.

There is much work to be done to develop hydrosols as a practical geoengineering options. For example, the potential utility of hydrosols will depend on the practical success of surface chemistry and engineering in overcoming and optimizing variable microbubble yields and lifetimes. As the ratio of the energy reflected to energy invested in microbubble deployment is potentially high ~ $10^4$ to $10^6$, to the extent that bubble lifetime extending surfactants can be found, extensive regional dispersal might increase ocean albedo enough to significantly reduce global net radiative forcing. Caution is needed, however, in evaluating this potential because the risks of albedo modification on phytoplankton ecology and biogeochemical cycles have yet to be studied (Wuebbles et al. 2001). Expanded oceanographic and limnological knowledge of natural surfactant and nanoparticle variability and the effect of hydrosols on $CO_2$ uptake (Le Quéré et. al 2007) is also needed to understand the optical properties and persistence of hydrosols, and to gauge both their environmental impacts and economic potential for conserving water in response to climate change (Milly et al. 2008).

It has been argued by Greene et al. (2010) that the growing effects of past emissions may be so severe as to render an exclusive focus on greenhouse gas regulation imprudent. While hydrosols pose both ecological and technical risks and challenges, their example is not so alien to the state of nature as using orbital mirrors or designer aerosols to alter the solar constant from the top down. In contrast, hydrosols have literally co-evolved with marine life (Lovelock 2009, Wurl 2008), and offer the possibility of managing albedo from the bottom up. As microbubbles already contribute to the albedo of the hydrosphere, using them to emulate nature on limited scales is less a matter of geoengineering than *geomimesis*, and should further research confirm the analysis offered here, they may offer a relatively benign way to conserve water while modulating climate change.

Such research is vital, for just as warming and water loss are inextricably linked; the historically recent phenomenon of fossil fuel use masks a far older and equally ominous problem. Mankind's albedo footprint began to grow in Neolithic times, and with the acceleration of demographic history, it has amplified ancient patterns of human land use to alter half the land surface of the Earth. (Pyne 1997, Seitz 2009, Ellis et al. 2010) Much of that albedo trend has been towards the dark side. Before we risk compounding the reflectivity loss from retreating sea ice and glaciers with geoengineering methods that may put the color of the sky at risk, we should seek new ways of lightening civilization's growing albedo footprint. Hydrosols may be one: to advance the synergy of local water conservation and global climate mitigation, don't dim the Sun. Brighten the water.

*Acknowledgements*: The author wishes to thank Zhonghai Jin for Mie model sensitivity runs, Kenneth Caldeira for GCM simulations, and Freeman Dyson, Richard Garwin, Isaac Silvera, Bruce Johnson, Marjorie Longo, Michael MacCracken, Arthur Rosenfeld, Howard Stone, Dariusz Stramski, Richard Wilson and Xiaodong Zhang for their helpful comments,the Riverforest Corporation for its donation of a microbubble generator and Harvard University for the appointment as Fellow in the Department of Physics that made the completion of this paper possible.

# References

Abkarian M, Subramaniam AB, Kim S-H,Larsn RJ, Yang S-M,and Stone H (2007) Dissolution arrest and stability of particle-covered bubbles. **Phys. Rev. Letters** 99: 188301-18830

Akbari S, Menon J and Rosenfeld A (2009) Global cooling: Increasing world-wide urban albedos to offset $CO_2$. **Climatic Change** 94:275–286. DOI 10.1007/s10584-008-9515-9

Brennen CE (2005) *Cavitation and Bubble Dynamics*, Oxford U. Press, New York

Buhaug Ø, Corbett JJ, Endresen Ø, Eyring V, Faber J, Hanayama S, Lee DS et al. 2009) *Second IMO GHG study* April 2009. International Maritime Organization (IMO) London

Crutzen P (2006) Albedo enhancement by stratospheric sulfur injections. **Climatic Change** 77: 211–219

D'Arrigo JS, Saiz-Jimemez C and Reimer NS (1984) Geochemical properties and biochemical composition of the surfactant mixture surrounding natural microbubbles in aqueous media. **J. Colloid and Interface Science** 100: 96-105

Dickey TD and Falkowski P (2003) Solar energy and its biological –physical interactions in the sea. In: A.R. Robinson et al. Eds. *The Sea*, V.12,; pp 401-4 John Wiley and Sons, New York

Dressaire E, Bee R, Bell C, Lips A and Stone HA Interfacial polygonal nanopatterning of stable microbubbles. **Science** 321:1198-1201

Ellis EC, Goldewijk KK, Siebert S, Lightman D and Ramankutty N (2010) Anthropogenic transformation of the biomes, 1700 to 2000 **Global Ecology and Biogeography** 19 ,5 :589–606 DOI: 10.1111/j.1466-8238.2010.00540

Gargett AE (1991) Physical processes and the maintenance of nutrient rich euphotic zones. **Limnol. Oceanogr.** 36: 1527- 1545

Gokbulak F and Ozban S (2006) Water loss through evaporation from water surfaces of lakes and reservoirs. **E-Water** http://www.ewaonline.de/journal/2006_07.pdf

Gordon HR, (1985) Ship perturbation of irradiance measurements at sea. 1: Monte Carlo simulations, **Appl. Opt**., 24: 4172–4182

Graham-Rove D (2008) Future Transport. **Nature** 454, 924-925

Greene CH, Baker D and Miller D(2010) A Very Inconvenient Truth **Oceanography** 23:214-221

Hoffman PF, Kaufman AJ, Halverson GP and Schrag DP, (1998) A Neoproterozoic snowball Earth. **Science**, 281:1341-1346

Hossein J, Syed J, Inoguchi Y, and Ma X (2010) *Surfactants,* http://www.sriconsulting.com/SCUP/Public/Reports/SURFA000s

Hwang PA, Xu D, and Wu J (1989) Breaking of wind-generated waves: Measurements and characteristics. **J. Fluid Mech**., 202, 177–200

Jin Z, Charlock T and Rutledge K (2002) Analysis of Broadband Solar Radiation and Albedo over the Ocean Surface at COVE **J. Atm. and Oceanic Tech.** 19 1585-1601

Jin Z ,Charlock TP, Rutledge K, Stamnes K and Wang Y (2006)Analytical solution of radiative transfer in the coupled atmosphere system with a rough surface;**Appl. Opt**. 45:7443-7455

Johnson BD Cooke RC (1980) Organic particle and aggregate formation resulting from the dissolution of bubbles in seawater. **Limnol. Oceanogr**. 25: 653-661

Johnson BD and Cooke RC (1981) Generation of Stabilized Microbubbles in Seawater. **Science** 213: 209-211

Johnson BD and Wangersky PJJ (1987) Microbubbles: Stabilization by Monolayers of Adsorbed Particles

*J. Geophys.Res.* 92, (C13):14641-14647

Kato H, (1999) *Skin Friction reduction by Microbubbles*. Toyo University Department of Engineering monograph, Pp 2-18 Kawagoe Japan

Katz ME, Wright JD, Miller BS, Cramer BS, Fennel K and Falkowski PG (2004) Evolutionary trajectories and biogeochemical impacts of marine eukaryotic phytoplankton *Ann. Rev. Ecol. Evol. Syst*. 35: 523-556

Keith DW, Ha-Duong M and Stolaroff JK (2005) Climate strategy with CO2 capture from the air *Climatic Change* 74 :17–45

Latham J (1990) Control of global warming? *Nature* 347: 339-340

Latham J (2002) Amelioration of Global Warming by Controlled Enhancement of the Albedo and Longevity of Low-Level Maritime Clouds *Atmos. Sci. Letters*. doi:10.1006/Asle.2002.0048

Latham J, Rasch P, Chen CC, Kettles L, Gadian A, Gettleman A, Morrison H, Bower K, and Choularton T. (2008) Global temperature stabilization via controlled albedo enhancement of low- level maritime clouds *Phil. Trans. R. Soc*. A 366: 3969–3987

Le Quéré C, Rödenbeck, Buitenhuis C, ConwayT, Langenfelds R, Gomez A, Labuschagne C, et al. (2007) Saturation of the Southern Ocean CO2 sink due to recent climate change. *Science*, 316: 1735-1738

Lozano MM, Talu E. and Longo ML (2007) Dissolution of microbubbles generated in seawater obtained offshore" Behavior and surface tension measurement *J. Geophys. Res*. 112, C:12001

Lozano MM and Longo ML: (2009) Microbubbles Coated with Disaturated Lipids and DSPPEG2000: Phase Behavior, Collapse Transitions, and Permeability, *Langmuir*, DOI: 10.1021/la803774q •.

Lovelock J (2009) A geophysiologist's thoughts on geoengineering *Phil. Trans. R. Soc*. A 366, 3883-3890 ; ; doi: 10.1098/rsta.2008.0135

MacCracken M (2009) On the possible use of geoengineering to moderate specific climate change impacts. *Environ. Res. Lett.* 4 :1-14

Mercado, LM, Bellouin N, Sitch S, Boucher O, Huntingford C, Wild M and Cox P (2009) Impact of changes in diffuse radiation on the global carbon land sink*. Nature* 458: 1014-17

Milly PCD, Betancourt J, Falkenmark M, Hirsch, RM, Kundzewicz ZW, Lettenmaier DP and Stouffer RJ (2008) Stationarity Is Dead: Whither Water Management?, *Science* 319: 573-574

Monahan ,EC and Mac Niocaill G ,Eds. (1986) *Oceanic Whitecaps and Their Role in Air-sea Exchange Processes:* pp 294 34- 55 ISBN 902772251X Springer Verlag Heidelberg

Moore KD, Voss KV and Gordon HR, (2000) Spectral reflectance of whitecaps: Their contribution to water-leaving radiance *J. Geophys. Res*. 105, C3: 6493-6499

Morton O (2009) Climate Crunch:Great white hope Nature 458:1097-1100 doi:10.1038/4581097a

Neiberger M (1957) Weather modification and smog *Science* 126: 637-645

Oleson K, Bonan GB and Feddema J (2010) Effects of white roofs on urban temperature in a global climate model. *Geophys. Res. Lett,* doi:10.1029/2009GL042194

Onhari H, Takahashi M and Himuro S (2002) Microbubble Generation in Sheared High Reynolds Number Flows *Japan*. *J. Multiphase Flow* 16:130-137

Ohnari H and Hiro J (2006) Micro and Nano Bubble Generation in Compressed Two Phase Water Jets *Japan. J. Multiphase Flow* 20, Pt.1:57-61

Pacala S and Socolow R (2004) Stabilization wedges: Solving the climate problem for the next 50 years with current technologies. *Science*, 305 (5686): 968–972

Piskozub J, Stramski D, Terrill, E and Melville WK (2009) Small-scale effects of underwater bubble clouds on ocean reflectance: 3-D modeling results. *Optics Express* 17:11747-11752

Pu G, Borden MA and Longo ML (2006) Collapse and Shedding Transitions in Binary Lipid Monolayers Coating Microbubbles *Langmuir*, 22, 2993-2999

Pyne S, Vestal Fire (1997) Pp 657 University of Washington Press, Seattle

Qin S, Caskey CF and Ferrara KW (2009) Ultrasound contrast microbubbles in imaging and therapy: physical principles and engineering *Phys Med Biol*. 54(6): R27.

Ramanathan V, Cess R, Harrison E, Minnis P, Barkstrom B, Ahmad E and Hartmann D (1989) Cloud-radiative forcing and climate: results from the earth radiation budget experiment *Science* 243: 57–69

Rasch PJ, Crutzen PJ, and Coleman DB (2008) Exploring the geoengineering of climate using stratospheric sulfate aerosols. *Geophys. Res. Lett.*, 35 L02809, doi:10.1029/2007GL032179

Read KA. Mahajan A Carpenter L, Evans MJ, Faria H, Saiz-Lopez A, Pilling MJ and Plane JMC (2009) Extensive halogen-mediated ozone destruction over the tropical Atlantic Ocean. *Nature* 453:1232-1235

Robock A, Marquardt AB, Kravitz B and Stenchikov G. (2009) Benefits, risks, and costs of stratospheric geoengineering *Geophys. Res. Lett.* 36 L:19703

Sadatomi M, Kawahara A, Matsuyama F and Kimura T (2007) An advanced microbubble generator and its application to a newly developed bubble-jet type air lift *Multiphase Sci. and Tech*.19:329-42

Salter S, Sortino G and Latham J(2008) Sea-going hardware for the cloud albedo method of reversing global warming. *Phil. Trans. R. Soc*. A 366 :3989–4006

Schneider SH (1996) Geoengineering: Could-or should- we do it? *Climatic Change* 33:291–302

Seitz F (1958) On the theory of the bubble chamber *Physics of Fluids* 1: 2-10

Seitz R (1986) Siberian fire as 'nuclear winter' guide; *Nature* 32, :116 - 117

Seitz R (1991) Black skies or pale fire? *Nature* 350: 182 - 183

Seitz R (2009) The next top model *Foreign Affairs* 88, 4 : 68

Shepherd J ed (2009) *Geoengineering the Climate: Science, Governance and Uncertainty* p 9 London Science Policy Centre, The Royal Society London

Solomon JS ed. Qin D, Manning M, Chen Z, Marquis M, Avery KB, Tignor M and Miller HL (2007) *Intergovernmental Panel on Climate Change Climate Change: The Physical Science Basis, Contribution of Working Group I to the Fourth Assessment Report of the IPCC*; p 996, Cambridge University Press, Cambridge

Stramski D and Tegowski J (2001) Effects of intermittent entrainment of air bubbles by breaking wind waves on ocean reflectance and underwater light field *J.Geophys.Res*.106,C12,31:345–360

Teller E, Wood L and Hyde R (1996) *Global warming and ice ages: I. Prospects for physics-based modulation of global change*.UCRL-JC-128715,Lawrence Livermore National Laboratory, Livermore, CA

Terrill E, Melville W.K. and Stramski D (2001) Bubble entrainment by breaking waves and their influence on optical scattering in the upper ocean *J.Geophys.Res.* 106, (C8): 16,815–16,823

Thorpe, SA (1982) The physics of breaking waves; *Philos.Trans. R.Soc*. A, 304: 155–210

Thorpe SA (1992) Bubble clouds and the dynamics of the upper ocean; *Q. J. R. Meteorol.Soc*.,118:1–22

Tilmes S, Müller R and Salawitch R (2008) The Sensitivity of Polar Ozone Depletion to Proposed Geoengineering Schemes; *Science* 320: 1201 - 1204

Toole DA, Siegel DA, Menzies DW, Neumann MJ, and Smith RC (2000) Remote-Sensing Reflectance Determinations in the Coastal Ocean Environment *Appl. Opt*., 39: 456–469

University Consortium on Atmospheric Research (2008) *Description of the NCAR Community Atmosphere Model (CAM3)*. http://www.cesm.ucar.edu/models/atm-cam/docs/description/

Van Vuuren DP, Eickhout B, Lucas PL, Meinshausen M, Plattner G-H, Joos F, Strassmann K, Smith S, Wigley T et al. (2009) Temperature increase of 21st century mitigation scenarios; *PNAS* 106 :9-16

Victor D, Morgan MG, Apt J, Steinbruner J and Ricke K (2009) The geoengineering option: A last resort against global warming? *Foreign Affairs* 88, 2 :64–76

Weber TC, Lyons AC and Bradley DL (2005) An estimate of the gas transfer rate from oceanic bubbles derived from multibeam sonar observations of a ship wake *J. Geophys. Res*. 110, C04005, doi:10.1029/2004JC002666

Willis J(1971) Some High Values for the Albedo of the Sea *J. Appl. Meteorology* 10: 1296-1392

Whitlock CH, Bartlett DS and Gurganus EA (1982) Sea foam reflectance and influence on optimum wavelength for remote sensing of ocean aerosols, *Geophys. Res. Lett.*, 9: 719–722

Wuebbles DJ, Naik V and Foley J (2001) Influence of geoengineered climate on the biosphere, *Eos*, 82:47

Wurl O and Holmes M (2008) The gelatinous nature of the sea- surface microlayer, *Marine Chemistry* 110: 89-97

Zhang X, Lewis M and Johnson B (1998) Influence of bubbles on scattering of light in the ocean; *Appl. Opt*., 37: 6525– 6536

Zhang X, Lewis M, Bissett WP, Johnson B and Kohler D (2004) Optical influence of ship wakes *Appl. Opt*, 43: 3122-3132

ADDITIONAL FIGURES

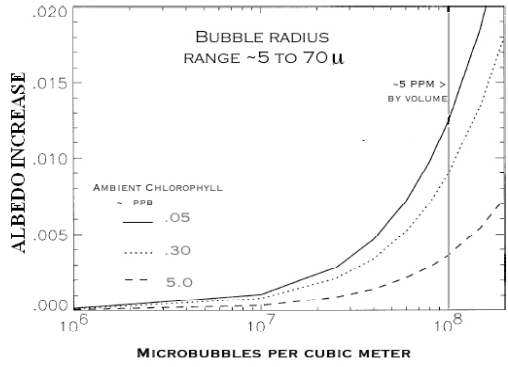

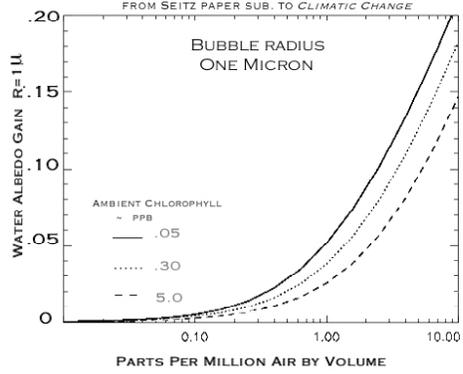

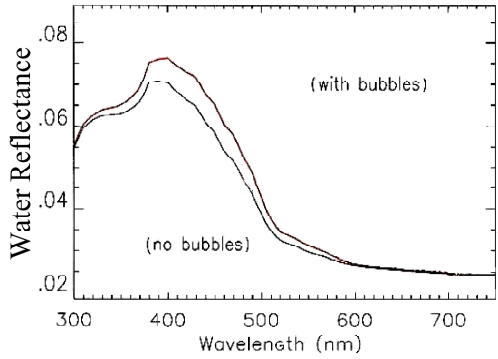

**Wavelength dependence of bubble reflectance**
(After Zhang 1998) at ~ 500 parts per billion by volume

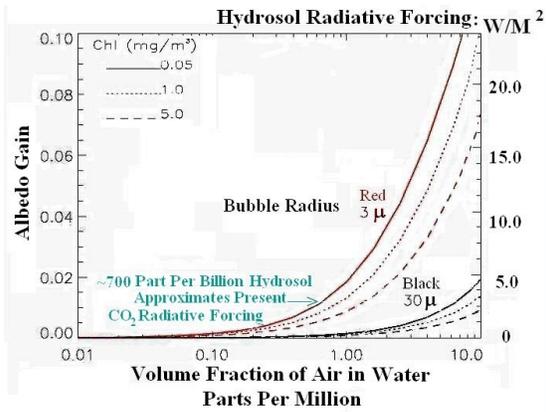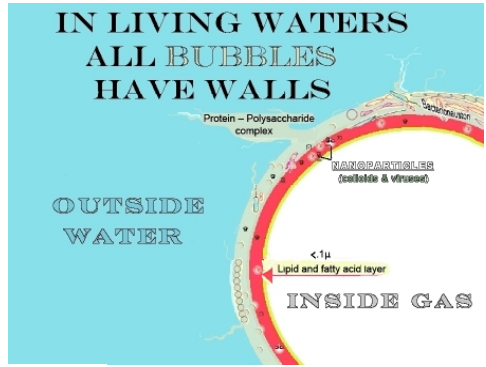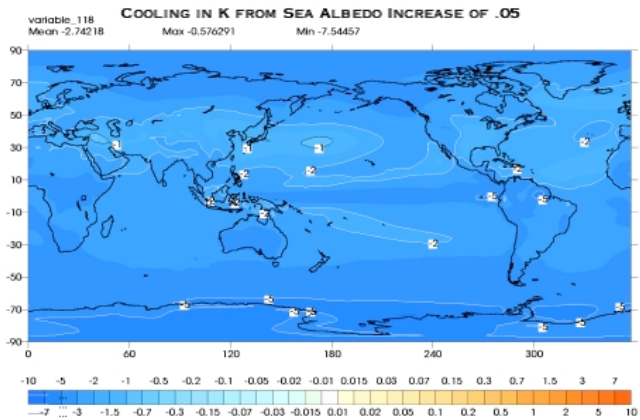